%% file: GHS-LLA-arxiv.tex
\documentclass[journal]{IEEEtran}

\ifCLASSINFOpdf
\else
\usepackage[dvips]{graphicx}
\fi
\usepackage[hidelinks]{hyperref}
\usepackage{url}

\usepackage{graphicx}
\usepackage{multicol}
\usepackage{amsmath}
\usepackage{amssymb}
\usepackage{amsthm}
\usepackage{caption}
\usepackage{xcolor}
\input{math-commands}
\usepackage{mathtools}
\newcommand{\sbt}{\,\begin{picture}(-1,1)(-1,-1)\circle*{3}\end{picture}\ }

\usepackage{algorithm}
\usepackage{algpseudocode}
\errorcontextlines\maxdimen
\makeatletter 
\newcommand*{\algrule}[1][\algorithmicindent]{\makebox[#1][l]{\hspace*{.5em}\vrule height .75\baselineskip depth .25\baselineskip}}%

\makeatletter
\g@addto@macro\normalsize{%
}

\newcount\ALG@printindent@tempcnta
\def\ALG@printindent{%
    \ifnum \theALG@nested>0
        \ifx\ALG@text\ALG@x@notext
            \addvspace{-3pt}
        \else
            \unskip
            \ALG@printindent@tempcnta=1
            \loop
                \algrule[\csname ALG@ind@\the\ALG@printindent@tempcnta\endcsname]%
                \advance \ALG@printindent@tempcnta 1
            \ifnum \ALG@printindent@tempcnta<\numexpr\theALG@nested+1\relax
            \repeat
        \fi
    \fi
    }%
\usepackage{etoolbox}
\patchcmd{\ALG@doentity}{\noindent\hskip\ALG@tlm}{\ALG@printindent}{}{\errmessage{failed to patch}}
\makeatother

\begin{document}
	
	\title{Maximum a Posteriori Estimation in Graphical Models Using Local Linear Approximation}
	
	\author{Ksheera Sagar, Jyotishka Datta, Sayantan Banerjee, Anindya Bhadra
		\thanks{Ksheera Sagar and Anindya Bhadra are with the Department of Statistics, Purdue University, IN 47907, USA (e-mails: kkeralap@purdue.edu, bhadra@purdue.edu).}
		\thanks{Jyotishka Datta is with the Department of Statistics, Virginia Tech, VA 24061, USA (email: jyotishka@vt.edu).}
		\thanks{Sayantan Banerjee is with the Operations Management and Quantitative Techniques Area, Indian Institute of Management Indore, MP 453556, India (email: sayantanb@iimidr.ac.in).}
    \thanks{The work of Anindya Bhadra was supported by
the U.S. National Science Foundation under Grant DMS-2014371.}}
	
	\maketitle
	
	\begin{abstract}
		Sparse structure learning in high-dimensional Gaussian graphical models is an important problem in multivariate statistical signal processing; since the sparsity pattern naturally encodes the conditional independence relationship among variables. However, maximum a posteriori (MAP) estimation is challenging under hierarchical prior models, and traditional numerical optimization routines or expectation--maximization algorithms are difficult to implement. To this end, our contribution is a novel local linear approximation scheme that circumvents this issue using a very simple computational algorithm. Most importantly, the condition under which our algorithm is guaranteed to converge to the MAP estimate is explicitly stated and is shown to cover a broad class of completely monotone priors, including the graphical horseshoe. Further, the resulting MAP estimate is shown to be sparse and consistent in the $\ell_2$-norm. Numerical results validate the speed, scalability, and statistical performance of the proposed method.
	\end{abstract}
	

	\IEEEpeerreviewmaketitle

	\section{Introduction}
	In multivariate statistical signal processing, precision matrix estimation is a potent tool for modeling conditional dependencies between random variables, with applications including sparse graph structure learning in Gaussian graphical models \cite{banerjee2015bayesian}, 
 channel estimation \cite{shariati2014low}, blind source separation \cite{ichir2005mean}, array signal processing \cite{hamza2019hybrid}, networked audio processing \cite{sim2006minimum}, and graph signal processing \cite{pavez2016generalized}, to name a few. The problem of precision matrix estimation is non-trivial in the high-dimensional regime, where the number of training samples is much lower than the number of variables; see, e.g., ~\cite{fan2016overview, senneret2016covariance, wei2021subspace}. 
 
 Let $Y_i \stackrel{i.i.d.} \sim \mathcal{N}(0,\Omega^{-1}), \; i=1,\ldots,n,$ be $n$ i.i.d. samples from a $q$-dimensional zero-mean Gaussian with  precision matrix $\Omega=\{\omega_{ij}\} \in \mathcal{M}_q^{+},$ the space of $q\times q$ positive definite matrices, and denote $Y=[Y_1,\ldots,Y_n]^{T} \in\mathbb{R}^{n\times q}$. The $q$ variables can also be represented as the nodes of an undirected graph equipped with a Gaussian distribution, resulting in a Gaussian graphical model (GGM), $G = (V,E)$, where $V$ is the set of $q$ nodes and $E$ is the edge-set. Via the Markov property, zero entries of $\Omega$ correspond to independence of the corresponding nodes conditioned on the rest \cite{lauritzen1996graphical}. Our goal is to obtain the \textit{maximum a posteriori} (MAP) estimate of $\Omega$ under a broad class of sparsity inducing priors within the Bayesian framework. Estimating a sparse precision matrix, $\Omega$, is also equivalent to graph structure learning, an important problem in network inference \cite{millington2019quantifying,lindsten2018graphical, dellaportas2003bayesian, van2022g}. However, implementing a traditional expectation-maximization (EM) algorithm for MAP estimation is difficult in hierarchical models or when the penalty, which is defined as the negative of the log prior density, is non-convex. In this paper, our contribution is a novel technique for MAP estimation in graphical models using the local linear approximation (LLA) algorithm \cite{zou2008one}, summarized in Algorithm~\ref{LLA_algo_pseudo_code}; and explicit identification of the conditions under which the LLA is guaranteed to find the MAP estimate, as derived in Theorem~\ref{prop:LLA-MAP-guarantee}. The required technique to establish our results lies in proving the complete monotonicity of the prior density on the elements of $\Omega$, which includes a large class of sparsity-inducing priors, including the graphical horseshoe \cite{li2019graphical}, on which our numerical results are based, but also priors such as the Laplace scale mixture of gamma~\cite{garrigues2010group}, special cases of power exponential scale mixtures~\cite{giri2016type,zhang2012ep} and the horseshoe-like distribution \cite{bhadra2021horseshoe}. Our approach is highly scalable compared to computation-intensive Markov chain Monte Carlo (MCMC) approaches, without sacrificing the statistical performance (see Tables~\ref{n120_p100_hubs} and~\ref{n120_p100_random}), producing results even when MCMC becomes computationally infeasible (see Supplementary Table S.II). Our approach also naturally results in a sparse estimate, unlike MCMC.
 
 \section{Background and Motivation}
	There is a rich Bayesian literature on precision matrix estimation in high dimensions; see \cite{banerjee2021bayesian}  for a review. Of these, the graphical horseshoe (GHS) model \cite{li2019graphical}  has proven to be extremely popular, with attractive theoretical and empirical properties; see \cite{shutta2022gaussian,williams2020back,jongerling2021bayesian}. The GHS model on $\Omega$ puts horseshoe priors \cite{carvalho2010horseshoe} on the off-diagonal elements $\omega_{ij}$ and an uninformative prior on the diagonal elements $\omega_{ii}$ to arrive at the following prior hierarchy for $i,j=1,\ldots,q$, subject to the constraint $\Omega \in \mathcal{M}_q^{+}$:
	\begin{eqnarray}
		\label{eqn:GHS-prior}
		\omega_{ij} \mid \lambda_{ij}, \tau  \stackrel{ind}\sim \NormRV(0, \lambda_{ij}^2\tau^2),\; i < j,\nonumber \\
		\lambda_{ij}  \stackrel{ind}\sim\CauchyRV^+(0,1),\; i< j, \; \omega_{ii} \propto 1,
	\end{eqnarray}
	where $\CauchyRV^+(0,1)$ is a half-Cauchy variable with density $p(x) \propto (1+x^2)^{-1},\; x \in \mathbb{R}^+$. Here $\lambda_{ij}$s are the local scale parameters, and $\tau > 0$ is a fixed unknown global scale parameter. 
 
Let us first outline some challenges with the prior hierarchy specified by \eqref{eqn:GHS-prior}. First, the marginal prior density $p(\omega_{ij} \mid \tau) = \int p(\omega_{ij} \mid \lambda_{ij}, \tau ) p(\lambda_{ij}) d\lambda_{ij} $ has no closed form in this model, and it is difficult to determine some basic properties of the marginal prior, such as whether it is convex. Similarly, it is difficult to arrive at an expectation--maximization (EM) routine for MAP estimation, due to a lack of closed form E-step updates. The only recourse is a computation-intensive fully Bayesian Markov chain Monte Carlo (MCMC) scheme as outlined by \cite{li2019graphical}, which is prohibitive in high dimensions.

To circumvent these issues while remaining theoretically sound, \cite{sagar2021precision} proposed a graphical horseshoe-like (GHS-like) prior for $\Omega$. The motivation behind the GHS-like prior is there is indeed a closed form to the marginal prior density on $\omega_{ij}$ in this model. Consequently, \cite{sagar2021precision} developed a successful EM scheme for MAP-estimation and established non-convexity properties of the GHS-like penalty. In this paper, we show it is possible to obtain MAP estimate under the original GHS model as well, even if the key barrier outlined above, i.e., the lack of closed form E-step updates in~\eqref{eqn:GHS-prior}, remains fully in place. The following remark clarifies the technical difficulties. 
\begin{remark}
It is worth noting here that the difficulty in implementing the usual EM algorithm for the horseshoe hierarchical model or its graphical version comes from the need to treat the local shrinkage parameters as latent variables, and not as tuning parameters that could be plugged in with their conditional modes. The latter approach \cite{lingjaerde2022scalable} evaluates $\argmax_{\Omega}p(\Omega \mid Y, \dagger)$ where $\dagger$ denotes the local modes for the other shrinkage parameters ($\lambda_{ij}$s in case of horseshoe) and not $\argmax_{\Omega}p(\Omega \mid Y)$, which is what we seek. Similarly, the desired E-step updates have no closed form. A cleaner and more principled workaround is to devise an LLA scheme using a mixture representation that avoids these potential pitfalls. An LLA approach to sparse precision matrix estimation also has the advantage of being portable to many other hierarchical shrinkage priors for which one might encounter the same challenges: a difficult EM due to a hierarchical model for the local shrinkage parameters, but a straightforward LLA for achieving the MAP. It is worth noting that the availability of full conditionals is sufficient for a Gibbs sampler or iterated conditional modes of \cite{besag1986statistical}, but not for EM.
\end{remark}
	
\section{Connection between LLA and MAP estimates of $\Omega$ under completely monotone priors }\label{sec:LLA}

The MAP estimate of $\Omega$ is defined as,
\begin{equation}
    \label{eq_log_likeli_and_penalty}
    \hat\Omega^{\mathrm{MAP}} = \underset{\Omega}{\text{argmin}}\left(\ell(\Omega;\,Y) + \text{pen}(|\Omega|)\right),
\end{equation}
where $\ell(\Omega;\,Y) = -(n/2)\left(\log\text{det}(\Omega) - \text{tr}(S\Omega)\right)$ is the negative log-likelihood, $S = n^{-1}Y^TY$ is the sample covariance, and $\text{pen}(|\Omega|)$ is the penalty function on $\Omega$ induced by the graphical shrinkage prior with penalty on individual elements given by $\text{pen}(|\omega_{ij}|)$. The LLA procedure approximates $\text{pen}(|\Omega|)$ with first order Taylor expansion of $\text{pen}(|\omega_{ij}|)$ as: 
\begin{equation*}
    \text{pen}(|\Omega|) \approx \underset{i\neq j}{\sum}\text{pen}\left(|\omega_{ij}^{(t)}|\right) +  \underset{i\neq j}{\sum}\text{pen}'\left(|\omega_{ij}^{(t)}|\right)(|\omega_{ij}|-|\omega_{ij}^{(t)}|),
\end{equation*}
where $\Omega^{(t)} = \{\omega_{ij}^{(t)}\}$ is the value of $\Omega$ at $t^\text{th}$ iteration of LLA and $\text{pen}'(|x|)$ denotes the first derivative of the penalty, $\text{pen}(|x|)$, with respect to $|x|$. Thus, the optimization problem in~\eqref{eq_log_likeli_and_penalty} leads to the iterative LLA update:
\begin{equation}
\label{omega_update_LLA}
    \Omega^{(t+1)} = \underset{\Omega}{\text{argmin}}\left(\ell(\Omega;\,Y) +  \underset{i\neq j}{\sum}\text{pen}'\left(|\omega_{ij}^{(t)}|\right)|\omega_{ij}|\right).
\end{equation}
To implement LLA, $G_{ij}^{(t)} := \text{pen}'\left(|\omega_{ij}|\right)$ at  $\omega_{ij}=\omega_{ij}^{(t)}$ needs to be computed, which is often feasible for mixture prior densities with no closed form. For example, see \cite[Appendix]{sagar2022laplace} for efficient computation of $G_{ij}^{(t)}$ under the horseshoe prior. Under the same prior, one can further establish $G_{ij}^{(t)}\rightarrow\infty$ as $\omega_{ij}^{(t)}\rightarrow 0$, proving \emph{local linear approximation of the penalty is the best convex majorization of the penalty function $\text{pen}(|\omega_{ij}|)$}, in the sense of ~\cite[Theorem 2]{zou2008one}.

The soft thresholding step in LLA, arising from the solution to the local $\ell_1$ penalized problem in~\eqref{omega_update_LLA}, naturally leads to a sparse estimate. On the contrary, the Bayes estimator of $\Omega$, obtained via drawing MCMC samples from the corresponding posterior distribution, as in \cite{li2019graphical}, is not sparse and is computationally burdensome, inherently limiting its scalability. 

While it is relatively easy to see that LLA results in a \emph{sparse} estimate, its relationship with the MAP estimate is not yet clear.  The following theorem explicitly identifies the condition under which the LLA procedure of~\eqref{omega_update_LLA} is guaranteed to lead to the MAP estimate of $\Omega$ defined in~ \eqref{eq_log_likeli_and_penalty}.

\begin{theorem}
\label{prop:LLA-MAP-guarantee}
    For a class of hierarchical graphical shrinkage priors in which the element-wise prior densities of the off-diagonal elements of the precision matrix $\Omega$ are completely monotone, the LLA algorithm is guaranteed to find the MAP estimate.
\end{theorem}
\begin{corollary}
\label{LLA-MAP-GHS}
The horseshoe density is completely monotone, and hence, the LLA procedure for the GHS prior of~\eqref{eqn:GHS-prior} produces the MAP estimate.
\end{corollary}
Proofs of Theorem~\ref{prop:LLA-MAP-guarantee} and Corollary~\ref{LLA-MAP-GHS} are in the Supplementary Material. Theorem~\ref{prop:LLA-MAP-guarantee} asserts our result is valid whenever the element-wise prior densities for the off-diagonal elements $\omega_{ij}$ are completely monotone. However, for a concrete illustration, we follow up on Corollary~\ref{LLA-MAP-GHS} and describe the LLA procedure for the graphical horseshoe model of~\eqref{eqn:GHS-prior} in detail.

\section{MAP Estimation procedure via LLA for GHS}
To update the elements of $\Omega$, we use the coordinate descent algorithm proposed by \cite{wang2012bayesian} for estimating covariance matrices under the graphical lasso prior. Without the loss of generality, we present detailed updates for elements of $\Omega$ in the $q^{th}$ column, and a pseudocode of the estimation procedure is detailed in Algorithm~\ref{LLA_algo_pseudo_code}. First, precision matrix $\Omega$ and the sample covariance matrix $S$ are partitioned into blocks as:
\begin{align}
	\Omega = \begin{bmatrix}
		\Omega_{11} & \Omega_{12}\\
		\Omega_{12}^T & \Omega_{22}\\
	\end{bmatrix}, \quad&nS = \begin{bmatrix}
		s_{11} & s_{12}\\
		s_{12}^T & s_{22}\\
	\end{bmatrix},\label{eq:Omega}
\end{align}
where $\Omega_{11}, s_{11}$ are $(q-1)\times(q-1)$ matrices comprising the first $(q-1)$ rows and columns of $\Omega, S$ respectively. Similarly, $\Omega_{12}, s_{12}$ are $(q-1)$ vectors comprising the off-diagonals in the $q^{th}$ columns of $\Omega, nS$; and $\Omega_{22}=\omega_{qq}$, a scalar. Define:
\begin{align*}
	\gamma  = \Omega_{22} - \Omega_{12}^T\Omega_{11}^{-1}\Omega_{12}, \quad& \beta  = \Omega_{12}. 
 \end{align*}
Then each term in~\eqref{omega_update_LLA} can be simplified as follows:
\begin{eqnarray}
	\log\text{det}(\Omega) &=& \log (\gamma) +c_1, \nonumber \\
	\text{tr}(nS\Omega) &=& 2s_{12}^T\beta + s_{22}\gamma + s_{22}\beta^T\Omega_{11}^{-1}\beta + c_2, \nonumber \\
	\underset{i\neq j}{\sum}G_{ij}^{(t)}|\omega_{ij}| &=& 2G_{\sbt\,q}^{(t)}|\beta| + c_3, \label{eq:G}
\end{eqnarray}
where $G_{\sbt\,q}^{(t)} = (G_{1p}^{(t)},\ldots, G_{q-1,\,q}^{(t)})$, $\beta=(\omega_{1q},\ldots,\omega_{q-1,\,q})^T$, $\abs{\beta} = \left(\abs{\omega_{1q}}, \ldots, \abs{\omega_{q-1,q}}\right)^T$ and  $c_1, c_2, c_3$ are independent of $\beta$ and $\gamma$. With these simplifications, the objective function in~\eqref{omega_update_LLA} in terms of $\beta,\,\gamma$ can be expressed as: 
\begin{equation}
\label{beta_gamma_obj_func}
	 -\frac{n}{2}\log (\gamma) + \frac{1}{2}\left[2s_{12}^T\beta + s_{22}\gamma + s_{22}\beta^T\left(\Omega_{11}^{(t)}\right)^{-1}\beta \right] +2G_{\sbt\,q}^{(t)}|\beta|.
\end{equation}
We can clearly see that minimizing the above function with respect to the entries of $\beta$ is analogous to solving a lasso problem. As we cannot update all entries of $\beta$ at once, we update each entry one at a time by conditional minimization. Without loss of generality, we outline the update for $\omega_{q-1,q}$. We now further partition the blocks above as follows:
\begin{equation*}	
\Omega_{11}^{-1} = \begin{bmatrix}
		C_{11} & C_{12}\\
		C_{12} & C_{22}\\
	\end{bmatrix},\, \beta =\left(\beta_{-(q-1)},\, \omega_{q-1,\,q}\right)^T,
\end{equation*}
where $\beta_{-(q-1)}= (\omega_{1q},\ldots, \omega_{q-2,q}),\; C_{11}$ is a matrix of dimension $(q-2)\times(q-2)$ which comprises the first $(q-2)$ rows and columns of the matrix $\Omega_{11}^{-1}$. Defining $s_{12}^{(q-1)}$ as the $(q-1)^\text{th}$ entry of $s_{12}$, the optimal value of  $\omega_{q-1,q}$ can be obtained by solving the lasso optimization problem,
\begin{eqnarray*}
	\omega_{q-1,q}^{(t+1)} &=\underset{\omega_{q-1,q}}{\text{argmin}}\left\lbrace s_{12}^{(q-1)}\omega_{q-1,q}+ 2G_{q-1,\,q}^{(t)}|\omega_{q-1,\,q}|+\right. \nonumber \\
	&  \left.	\frac{s_{22}}{2}\left(2C_{12}^{(t)T}\beta_{-(q-1)}^{(t)}\omega_{q-1,\,q}+ C_{22}^{(t)}\omega_{q-1,\,q}^2  \right) \right\rbrace .
\end{eqnarray*}
Denoting $\hat{\omega}^{(t)}_{q-1,\,q} = \left(s_{22}C_{22}^{(t)}\right)^{-1}\left(s_{12}^{(q-1)} + s_{22}C_{12}^{(t)T}\beta_{-(q-1)}^{(t)}\right)$ and $\eta_{q-1,\,q}^{(t)} = 2\left(s_{22}C_{22}^{(t)}\right)^{-1}G_{q-1,\,q}^{(t)}$, we get:
\begin{equation*}
	\omega_{q-1,q}^{(t+1)} = 
	\begin{cases}
		-\eta_{q-1,\,q}^{(t)}  - \hat{\omega}^{(t)}_{q-1,\,q},& \text{if }\text{ } \hat{\omega}^{(t)}_{q-1,\,q} < -\eta_{q-1,\,q}^{(t)} ,\\
		\eta_{q-1,\,q}^{(t)}  - \hat{\omega}^{(t)}_{q-1,\,q},& \text{if }\text{ } \hat{\omega}^{(t)}_{q-1,\,q} >\eta_{q-1,\,q}^{(t)} ,\\
		0, & \text{if }\text{ } |\hat{\omega}^{(t)}_{q-1,\,q}| < \eta_{q-1,\,q}^{(t)} .
	\end{cases}
\end{equation*}
Similarly updating all entries of $\beta$ and minimizing~\eqref{beta_gamma_obj_func} over $\gamma$, we have the final updates of $\gamma$ and $\beta$ as: 
\begin{equation}
\gamma^{(t+1)} = n/s_{22},\;\; \beta^{(t+1)} =\left(\omega_{1,q}^{(t+1)},\ldots, \omega_{q-1,q}^{(t+1)}\right)^T. \label{eq:bg}
\end{equation}
With this, the final updates of the $q^\text{th}$ column of $\Omega$, for $(t+1)^{th}$ iteration becomes, 
\begin{eqnarray}
\label{column_updates_in_Omega}
\Omega_{12}^{(t+1)} &=&\beta^{(t+1)}, \nonumber \\
\Omega_{22}^{(t+1)} &=& \gamma^{(t+1)} + \beta^{(t+1)T}\left(\Omega_{11}^{(t)}\right)^{-1}\beta^{(t+1)}.    
\end{eqnarray}
Cycling through all columns  $q-1\rightarrow 1$, gives the final update $\Omega^{(t)}\rightarrow\Omega^{(t+1)}$, completing the description. Our update is guaranteed to maintain the positive definiteness of $\Omega^{(t)}$ at every iteration, see~\cite{wang2014coordinate}. The complete procedure is summarized in Algorithm~\ref{LLA_algo_pseudo_code} with computer code  publicly available at: \href{https://github.com/sagarknk/GHS-LLA-codes}{https://github.com/sagarknk/GHS-LLA-codes}. Finally, we note that the only step specific to GHS in the entire procedure is the computation of the derivative of the horseshoe penalty, $G_{ij}^{(t)}$, in~\eqref{eq:G} following \cite[Appendix]{sagar2022laplace}, and the algorithm applies identically to penalties induced by other completely monotone densities, with this term suitably modified.

\begin{algorithm}[!t]
		\caption{LLA algorithm for MAP estimation of $\Omega$}\label{LLA_algo_pseudo_code}
   \hspace*{\algorithmicindent} \textbf{Input: } \text{Initial value} $\Omega_{\mathrm{init}}$; Sample covariance $S$; sample size $n$; dimension $q$; algorithm tolerance=$\mathrm{tol}$.\\
 \hspace*{\algorithmicindent} \textbf{Output: }$\hat\Omega^{\mathrm{MAP}}$, a local optimum solution of~\eqref{eq_log_likeli_and_penalty}.\\
 \hspace*{\algorithmicindent} \textbf{Define: }$\text{PER}(Z;\, i, j)$: a subroutine to simultaneously swap row and column $i$ and $j$ of a symmetric matrix $Z$. 
		\begin{algorithmic}
			\Function{Graphical\_MAP}{$\Omega_{\mathrm{init}}, S, n, q, \mathrm{tol}$}\vspace{0.1cm} 
            \State Initialize $t\leftarrow 0,\; \Omega^{(t)} \leftarrow \Omega_{\mathrm{init}}, \; \Delta > \mathrm{tol}$.
			\While{$\Delta \geq \text{tol}$}
   \State Save $\Omega_{\mathrm{current}} \leftarrow \Omega^{(t)}$.
   \State \small\texttt{// For loop to update row/cols. of $\Omega$.}\normalsize
			\For{$i = \{q, \ldots,1\}$}
   \State \small\texttt{// Swap $i$th and $q$th row/col.}\normalsize
                \State Set \small$\Omega^{(t)} \leftarrow \text{PER}(\Omega^{(t)};\, i, q)$, $S\leftarrow \text{PER}(S;\, i, q)$.\normalsize             
  \State \small\texttt{// Optimize over $q$th row/col.}\normalsize
  \State Update $\beta^{(t+1)}$ and $\gamma^{(t+1)}$ according to \eqref{eq:bg}.
            \State Set $\Omega_{12}^{(t+1)},\,\Omega_{22}^{(t+1)}$ according to~\eqref{column_updates_in_Omega}.
            \State Form $\Omega^{(t)}$ using $\Omega_{11}^{(t)},\,\Omega_{12}^{(t+1)},\,\Omega_{22}^{(t+1)}$ by \eqref{eq:Omega}.
                         \State \small\texttt{// Swap back $q$th and $i$th row/col.}\normalsize
            \State Set  \small$\Omega^{(t)} \leftarrow \text{PER}(\Omega^{(t)};\, i, q),\, S \leftarrow \text{PER}(S;\, i, q)$.\normalsize
			\EndFor
   \State Set $\Omega^{(t+1)}\leftarrow \Omega^{(t)}$. \small\texttt{// $\Omega$ update complete.}
            \State Update $\Delta  \leftarrow |\!|\Omega^{(t+1)} - \Omega_{\mathrm{current}}|\!|_2; \; t\leftarrow t+1$.
			\EndWhile
			\State \Return $\hat\Omega^{\mathrm{MAP}} \leftarrow \Omega^{(t)}$.
			\EndFunction
		\end{algorithmic}
	\end{algorithm}

\section{Consistency of the MAP estimator}

In the previous sections, we developed an LLA procedure and established the required conditions for LLA to find the MAP estimate. We now present our result on the consistency of the MAP estimate of $\Omega$ under the GHS prior in the $\ell_2$-norm $\|\cdot\|_2$, defined as $\|A\|_2 = \sqrt{\text{trace}(A^T A)}$. The assumptions required for this result are identical to that required for establishing convergence rates of the posterior distribution of $\Omega$ using the GHS prior, as outlined in \cite{sagar2021precision}; see Assumptions~4.1--4.5 and Corollary~4.7 therein.

\begin{theorem}
\label{th:map}
The MAP estimator of $\Omega$ under GHS, denoted by $\hat{\Omega}^{\mathrm{MAP}}$, is consistent, in the sense that as $n\rightarrow\infty$, the $\ell_2$-norm of the difference between the MAP estimator and the true precision matrix $\Omega_0$ converges to zero in probability, that is,
\small
\begin{equation*}
    \|\hat{\Omega}^{\mathrm{MAP}} - \Omega_0\|_2 \stackrel{P}{\rightarrow} 0,
\end{equation*}
\normalsize
with convergence rate given by $\epsilon_n = n^{-1/2}(q+s)^{1/2}(\log q)^{1/2}$, where $s$ is the number of non-zero off-diagonal elements in $\Omega_0$. 
\end{theorem}
Proof of Theorem~\ref{th:map} is in the Supplementary Material. This result establishes that the MAP estimator itself converges to the true precision matrix when the sample size is large. A key step lies in proving that the penalty function induced by the graphical horseshoe has bounded first and second derivatives, which we prove, while appealing to Theorem 4.9 of~\cite{sagar2021precision} to complete the rest of the proof.

\section{Numerical results}
In this section we compare the MAP estimates obtained using the local linear approximation (LLA) of graphical horseshoe penalty under Laplace (LLA (l)) and half-Cauchy (LLA (c)) mixture representations (see Proposition 1 and (1) of \cite{sagar2022laplace}), against the MCMC estimate under the graphical horseshoe (GHS) prior~\cite{li2019graphical} and frequentist graphical lasso~\cite{friedman2008sparse}, with penalized (GL1) and unpenalized (GL2) diagonal elements. All the horseshoe methods are implemented in MATLAB and the graphical lasso is implemented in R-package \texttt{glasso}~\cite{R:glasso}. For simulations, we consider two problem dimensions $(n,q)=\{(120,100),\,(120,200)\}$ and two structures of precision matrix: `hubs',\,`random'~(see \cite{friedman2008sparse, li2019graphical, sagar2021precision} for more details on the precision matrix structure). Given $(n,q)$ and the true precision matrix $\Omega_0$, we generate 50 data sets from $\mathcal{N}(0,\Omega_0^{-1})$ and estimate the precision matrix $\hat{\Omega}$ using the methods mentioned above. Tuning parameters for graphical lasso and LLA based methods are chosen by 5-fold cross validation; see~\cite[Appendix]{sagar2022laplace} for other ways of tuning the global scale parameter $\tau$. The stopping criterion for LLA based methods is set as $|\!|\Omega^{(t+1)} - \Omega^{(t)}|\!|_2<10^{-3}$ and we draw a total of 6000 samples from the posterior (with 1000 burn-in samples), for the GHS estimate. Further, the estimate for LLA based methods is computed by taking the average of estimates obtained using 50 randomly generated start points i.e, estimating MAP using Algorithm~\ref{LLA_algo_pseudo_code} with 50 different start points ($\Omega^{(0)}$) and taking the average of resulting $\hat{\Omega}^\text{MAP}$. The use of randomly chosen starting values is a typical strategy to reduce the sensitivity to starting points for a local optimization algorithm such as EM or coordinate-descent ~\cite{sagar2021precision}. 

We compare the estimates using Stein's loss (=$\text{tr}(\hat{\Omega}\Omega_0^{-1}) -\log|\hat{\Omega}\Omega_0^{-1}| -q$), a measure of empirical KL divergence, Frobenius norm (F norm = $\|\hat{\Omega} - \Omega_0^{-1}\|_2$), true and false positive rates of detecting the non-zero signals (TPR, FPR respectively) in $\hat{\Omega}$ and Matthews Correlation Coefficient (MCC). The mean (sd) of these performance metrics of precision matrix estimates for $(n,q) = (120, 100)$, over 50 generated data sets, is presented in Tables~\ref{n120_p100_hubs}--\ref{n120_p100_random}. The results for $(n,q) = (120, 200)$ and $(n,q) = (300, 850)$ are presented in the Supplementary Material. In point estimation methods, one can get exact zeros in the off-diagonal entries of the precision matrix estimate, which makes detection of signals straightforward. But for Bayesian procedures (like GHS), we use the middle 50\% posterior credible intervals for signal identification~\cite{li2019graphical}.  

From the results in Tables~\ref{n120_p100_hubs}--\ref{n120_p100_random} and Tables~S.I--S.II, it is clear that the estimates LLA (l) and LLA (c) perform similarly. In general, they also have the lowest or comparable  Stein's loss, F norm and the second best FPR, MCC among the competing procedures. It can also be seen that the LLA based methods are computationally several orders of magnitude faster than fully Bayesian MCMC of \cite{li2019graphical}, and result in better statistical estimates compared to the graphical lasso.

\begin{table}[!htb]
\centering
\caption{Comparison of results for competing procedures. $n=120,\, q=100$ and Hubs structure.}
\label{n120_p100_hubs}
\begin{tabular}{|c|ccccc|}
\hline
& LLA (l) & LLA (c) & GHS & GL1 & GL2 \\ \hline
Stein's loss      & \begin{tabular}[c]{@{}c@{}}  3.873\\ (0.379)\end{tabular}   & \begin{tabular}[c]{@{}c@{}}3.898\\ (0.389)\end{tabular}  &\begin{tabular}[c]{@{}c@{}} 5.1 \\ (0.454)\end{tabular}  & \begin{tabular}[c]{@{}c@{}} 5.255 \\ (0.263)\end{tabular}  & \begin{tabular}[c]{@{}c@{}} 6.328 \\ (0.414)\end{tabular}   \\
F norm       & \begin{tabular}[c]{@{}c@{}}  2.235\\ (0.099)\end{tabular}   & \begin{tabular}[c]{@{}c@{}}2.246  \\ (0.098)\end{tabular}  &\begin{tabular}[c]{@{}c@{}}  2.547\\ (0.128)\end{tabular}  & \begin{tabular}[c]{@{}c@{}}  3.018\\ (0.091)\end{tabular}  & \begin{tabular}[c]{@{}c@{}}  3.432\\ (0.112)\end{tabular}   \\
TPR      & \begin{tabular}[c]{@{}c@{}}  0.967\\ (0.024)\end{tabular}   & \begin{tabular}[c]{@{}c@{}} 0.967 \\ (0.025)\end{tabular}  &\begin{tabular}[c]{@{}c@{}} 0.871 \\ (0.04)\end{tabular}  & \begin{tabular}[c]{@{}c@{}} 0.995 \\ (0.007)\end{tabular}  & \begin{tabular}[c]{@{}c@{}} 0.986\\(0.017)\end{tabular}   \\
FPR       & \begin{tabular}[c]{@{}c@{}} 0.032 \\ (0.012)\end{tabular}   & \begin{tabular}[c]{@{}c@{}}  0.032  \\ (0.012)\end{tabular}  &\begin{tabular}[c]{@{}c@{}}  0.003\\ (0.001)\end{tabular}  & \begin{tabular}[c]{@{}c@{}} 0.101 \\ (0.016)\end{tabular}  & \begin{tabular}[c]{@{}c@{}} 0.045 \\ (0.008)\end{tabular}   \\
MCC       & \begin{tabular}[c]{@{}c@{}} 0.592 \\ (0.063)\end{tabular}   & \begin{tabular}[c]{@{}c@{}}  0.593 \\ (0.064)\end{tabular}  &\begin{tabular}[c]{@{}c@{}}  0.848\\ (0.028)\end{tabular}  & \begin{tabular}[c]{@{}c@{}}  0.373\\ (0.027)\end{tabular}  & \begin{tabular}[c]{@{}c@{}} 0.523 \\ (0.0391)\end{tabular}   \\
Time (s) & 3.04 & 3.09 & 250.12 & 1.59 & 1.65 \\ \hline
\end{tabular}
\end{table}
\vspace{-0.5cm}
\begin{table}[!htb]
\centering
\caption{Comparison of results for competing procedures. $n=120,\, q=100$ and Random structure.}
\label{n120_p100_random}
\begin{tabular}{|c|ccccc|}
\hline
& LLA (l) & LLA (c) & GHS & GL1 & GL2 \\ \hline
Stein's loss      & \begin{tabular}[c]{@{}c@{}} 2.432 \\ (0.296)\end{tabular}   & \begin{tabular}[c]{@{}c@{}} 2.428 \\ (0.288)\end{tabular}  &\begin{tabular}[c]{@{}c@{}}  2.173 \\ (0.28)\end{tabular}  & \begin{tabular}[c]{@{}c@{}}  5.245\\ (0.254)\end{tabular}  & \begin{tabular}[c]{@{}c@{}} 6.785 \\ (0.464)\end{tabular}   \\
F norm       & \begin{tabular}[c]{@{}c@{}} 1.997 \\ (0.132)\end{tabular}   & \begin{tabular}[c]{@{}c@{}} 1.981 \\ (0.132)\end{tabular}  &\begin{tabular}[c]{@{}c@{}}  1.961 \\ (0.144)\end{tabular}  & \begin{tabular}[c]{@{}c@{}}  3.348\\ (0.115)\end{tabular}  & \begin{tabular}[c]{@{}c@{}}  4.084\\ (0.143)\end{tabular}   \\
TPR      & \begin{tabular}[c]{@{}c@{}} 0.874 \\ (0.05)\end{tabular}   & \begin{tabular}[c]{@{}c@{}}  0.881\\ (0.049)\end{tabular}  &\begin{tabular}[c]{@{}c@{}}  0.82 \\ (0.043)\end{tabular}  & \begin{tabular}[c]{@{}c@{}}  0.951\\ (0.03)\end{tabular}  & \begin{tabular}[c]{@{}c@{}}  0.882\\ (0.038)\end{tabular}   \\
FPR       & \begin{tabular}[c]{@{}c@{}}  0.019\\ (0.009)\end{tabular}   & \begin{tabular}[c]{@{}c@{}}0.021  \\ (0.009)\end{tabular}  &\begin{tabular}[c]{@{}c@{}} 0.0005 \\ (0.0003)\end{tabular}  & \begin{tabular}[c]{@{}c@{}}  0.101\\ (0.013)\end{tabular}  & \begin{tabular}[c]{@{}c@{}}0.045  \\ (0.007)\end{tabular}   \\
MCC       & \begin{tabular}[c]{@{}c@{}} 0.481 \\ (0.078)\end{tabular}   & \begin{tabular}[c]{@{}c@{}}0.464  \\ (0.068)\end{tabular}  &\begin{tabular}[c]{@{}c@{}}  0.868\\ (0.032)\end{tabular}  & \begin{tabular}[c]{@{}c@{}}  0.232 \\ (0.018)\end{tabular}  & \begin{tabular}[c]{@{}c@{}}  0.321 \\ (0.024)\end{tabular}   \\
Time (s) & 6.53 & 6.56 & 253.41 & 4.14  &  4.61\\ \hline
\end{tabular}
\end{table}

\section{Concluding remarks}

We devise a novel MAP estimation scheme in GGMs under a general class of completely monotone priors, including the GHS. The algorithmic procedure is an LLA scheme, and equivalence between LLA and MAP estimates is established by first showing the complete monotonicity of the prior density, and then appealing to Theorem 3 of \cite{zou2008one}. The resultant estimate is naturally sparse, owing to the soft thresholding step in LLA, aiding graph structure learning. This is at a contrast with the fully Bayesian estimate, which requires some form of post-processing of the MCMC samples, such as thresholding, to yield a sparse estimate. MAP estimation via LLA is not restricted to the graphical horseshoe prior model as outlined in this paper, but can be used for a broad class of hierarchical shrinkage prior models for precision matrix estimation, where the element-wise prior densities are complete monotone and the induced penalty is strongly concave, for example, the family of Laplacian scale mixture densities~\cite{garrigues2010group, polson2014bayesian,armagan2013generalized}. Implementing a traditional EM algorithm becomes non-trivial in such scenarios owing to the local shrinkage parameters in the hierarchical representation, whereas an LLA algorithm avoids such bottlenecks. Our work has thus reduced the difficult problem of evaluating a consistent MAP estimator under a general class of graphical shrinkage priors, to  establishing complete monotonicity of the prior. This technique can be extended to structure learning of multiple graphs with a shared structure; a problem we propose as future work. 

\clearpage
\bibliographystyle{IEEEtran}
\IEEEtriggeratref{18}
\bibliography{hs-review}

\setcounter{table}{0}
\setcounter{section}{0}
\setcounter{equation}{0}
\renewcommand\thetable{S.\Roman{table}}
\renewcommand\thesection{S.\Roman{section}}

\onecolumn
\begin{center}
	{\huge{Supplementary Material}} 
\end{center}

\begin{multicols}{2}
\section{Proof of Theorem 2 and Corollary 3}
 If the element-wise prior densities $p(\omega_{ij})$ are completely monotone, the induced penalty function admits a completely monotone derivative \cite[Theorem 4.1.5]{bochner1955harmonic}. Then, via Theorem~3 in \cite{zou2008one}, for such completely monotone prior densities,  the LLA is equivalent to the EM algorithm, and hence is guaranteed to produce the MAP estimate. This completes the Proof of Theorem 2.

 Since the horseshoe density can also be expressed as a Laplace scale mixture \cite[Proposition 1]{sagar2022laplace}, the density is completely monotone, via the Bernstein--Widder theorem \cite[Theorem 4.1.1]{widder2015laplace}. This completes the proof of Corollary 3.

\section{Proof of Theorem 4}

The Proof of Theorem 4 follows closely the proof of Theorem 4.9 in~\cite{sagar2021precision}, under identical assumptions. Yet, this proof requires a key step to be proven, which is the boundedness of the first and second derivatives of the horseshoe penalty. Plugging these bounds into Equation (22) of~\cite{sagar2021precision}, the rest of the proof follows. To establish these bounds, we begin with the bounds of the horseshoe density, $p_{HS}(|x|)$, a result from \cite[Proof of Theorem 1]{carvalho2009handling}, using which the derivative of the horseshoe penalty (with $\tau=1$), $\text{pen}'\left(|x|\right) =- d/d|x|(\log p_{HS}(|x|))$ can bounded as,
\small
\[
\frac{2}{|x| \log\left(1 + \dfrac{2}{|x|^2}\right)} - |x| < 
\text{pen}'(|x|) < 
\frac{4}{|x| \log\left(1 + \dfrac{4}{|x|^2}\right)}- |x|.
\]
\normalsize
Note that, $\text{pen}'(|x|)$ is completely monotone~\cite{sagar2022laplace}, and hence non-negative. Thus, the right-hand side of the above bound will suffice to bound $|\text{pen}'(|x|)|.$ The right-hand side can be written as:
\small
\begin{eqnarray}
&&4\left(1 + \dfrac{4}{|x|^2}\right)\left\{x \left(1 + \dfrac{4}{|x|^2}\right)\log\left(1 + \dfrac{4}{|x|^2}\right)\right\}^{-1} - |x| \nonumber \\
&<& 4\left(1 + \dfrac{4}{|x|^2}\right)\left\{|x| \cdot\dfrac{4}{|x|^2}\right\}^{-1} - |x| =  \dfrac{4}{|x|}. \nonumber
\end{eqnarray}
\normalsize
The inequality above follows from the fact that $(1+|x|)\log(1+|x|) > |x|$. Hence, $|\text{pen}'(|x|)| < 4/|x|.$ The second derivative of the penalty function, $\text{pen}''(|x|)$, is given by:
\small
\begin{eqnarray}
&& -\text{pen}''(|x|)p^{-1}(|x|) + \{\text{pen}'(|x|)p^{-1}(|x|)\}^{2} \nonumber \\
&=& -1 + |x| \text{pen}'(|x|) - 2K|x|^{-2}p^{-1}(|x|) + \{\text{pen}'(|x|)\}^2 \nonumber \\
&<& |x| \text{pen}'(|x|) + \{\text{pen}'(|x|)\}^2,\,\mathrm{since}\, 1 + 2K|x|^{-2}p^{-1}(|x|) > 0. \nonumber
\end{eqnarray}
\normalsize
Hence, 
\[
|\text{pen}''(|x|)| < |x|\dfrac{4}{|x|} + \dfrac{16}{|x|^2} = 4\left(1 + \dfrac{4}{|x|^2}\right).
\]

\section{Additional numerical results}
We present results for the competing procedures for precision matrix estimation corresponding to $(n,q)=(120,200)\text{ and }(n,q)=(300,850)$ for the `hubs' structure in the tables below. In Table~\ref{n300_p850_random} results for the fully Bayesian procedure (GHS) are not available because for this $q$  it failed to complete the same number of MCMC iterations as in the other tables in 24 hours.
\begin{center} 
		\captionof{table}{Comparison of results for competing procedures. $n=120,\, q=200$ and Hubs structure.}\label{n120_p200_hubs}
		\small 
\begin{tabular}{|c|ccccc|}
\hline
& LLA (l) & LLA (c) & GHS & GL1 & GL2 \\ \hline
Stein's loss      & \begin{tabular}[c]{@{}c@{}} 8.835  \\ (0.672)\end{tabular}   & \begin{tabular}[c]{@{}c@{}} 8.879 \\ (0.687)\end{tabular}  &\begin{tabular}[c]{@{}c@{}}  11.67 \\ (0.763)\end{tabular}  & \begin{tabular}[c]{@{}c@{}} 12.407 \\ (0.491)\end{tabular}  & \begin{tabular}[c]{@{}c@{}}  15.243\\ (0.819)\end{tabular}   \\
F norm       & \begin{tabular}[c]{@{}c@{}} 3.306 \\ (0.114)\end{tabular}   & \begin{tabular}[c]{@{}c@{}} 3.325 \\ (0.115)\end{tabular}  &\begin{tabular}[c]{@{}c@{}} 3.752 \\ (0.131)\end{tabular}  & \begin{tabular}[c]{@{}c@{}} 4.594 \\ (0.1)\end{tabular}  & \begin{tabular}[c]{@{}c@{}} 5.298 \\ (0.152)\end{tabular}   \\
TPR      & \begin{tabular}[c]{@{}c@{}} 0.949 \\ (0.024)\end{tabular}   & \begin{tabular}[c]{@{}c@{}} 0.95 \\ (0.024)\end{tabular}  &\begin{tabular}[c]{@{}c@{}}0.775  \\ (0.0.32)\end{tabular}  & \begin{tabular}[c]{@{}c@{}} 0.99 \\ (0.007)\end{tabular}  & \begin{tabular}[c]{@{}c@{}} 0.976 \\ (0.014)\end{tabular}   \\
FPR       & \begin{tabular}[c]{@{}c@{}} 0.015 \\  (0.005)\end{tabular}   & \begin{tabular}[c]{@{}c@{}} 0.015 \\ (0.005)\end{tabular}  &\begin{tabular}[c]{@{}c@{}} 0.001 \\ (0.0002)\end{tabular}  & \begin{tabular}[c]{@{}c@{}} 0.065 \\ (0.005)\end{tabular}  & \begin{tabular}[c]{@{}c@{}} 0.024 \\ (0.006)\end{tabular}   \\
MCC       & \begin{tabular}[c]{@{}c@{}} 0.595 \\ (0.053)\end{tabular}   & \begin{tabular}[c]{@{}c@{}}  0.593 \\ (0.056)\end{tabular}  &\begin{tabular}[c]{@{}c@{}} 0.817 \\ (0.025)\end{tabular}  & \begin{tabular}[c]{@{}c@{}}  0.336\\ (0.015)\end{tabular}  & \begin{tabular}[c]{@{}c@{}}  0.515\\ (0.043)\end{tabular}   \\
Time (s) & 17.73 & 18.93 & 1865.69 & 13.35 &  14.92\\ \hline
\end{tabular}
\end{center}

\begin{center} 
		\captionof{table}{Comparison of results for competing procedures. $n=300,\, q=850$ and Hubs structure.}\label{n300_p850_random}
		\small 
\begin{tabular}{|c|ccccc|}
\hline
& LLA (l) & LLA (c) & GHS & GL1 & GL2 \\ \hline
Stein's loss      & \begin{tabular}[c]{@{}c@{}} 39.924 \\ (0.471)\end{tabular}   & \begin{tabular}[c]{@{}c@{}} 37.741\\ (0.574)\end{tabular}  & - & \begin{tabular}[c]{@{}c@{}}  34.643\\ (0.477)\end{tabular}  & \begin{tabular}[c]{@{}c@{}} 41.378 \\ (0.513)\end{tabular}   \\
F norm       & \begin{tabular}[c]{@{}c@{}}  7.085\\ (0.021)\end{tabular}   & \begin{tabular}[c]{@{}c@{}} 6.887 \\ (0.011)\end{tabular}  & -  & \begin{tabular}[c]{@{}c@{}}  8.098\\ (0.051)\end{tabular}  & \begin{tabular}[c]{@{}c@{}}  9.176\\ (0.048)\end{tabular}   \\
TPR      & \begin{tabular}[c]{@{}c@{}}  0.969 \\ (0.01)\end{tabular}   & \begin{tabular}[c]{@{}c@{}} 0.977 \\ (0.009)\end{tabular}  & - & \begin{tabular}[c]{@{}c@{}}  1\\ (0)\end{tabular}  & \begin{tabular}[c]{@{}c@{}} 1 \\ (0.0003)\end{tabular}   \\
FPR       & \begin{tabular}[c]{@{}c@{}} 0.00004 \\ (0.00002)\end{tabular}   & \begin{tabular}[c]{@{}c@{}} 0.0001  \\ (0.00001)\end{tabular}  & - & \begin{tabular}[c]{@{}c@{}}  0.023\\ (0.0005)\end{tabular}  & \begin{tabular}[c]{@{}c@{}} 0.008 \\ (0.0003)\end{tabular}   \\
MCC       & \begin{tabular}[c]{@{}c@{}}  0.975 \\ (0.009)\end{tabular}   & \begin{tabular}[c]{@{}c@{}}  0.976 \\ (0.006)\end{tabular}  & -   & \begin{tabular}[c]{@{}c@{}}0.288 \\ (0.003)\end{tabular}  & \begin{tabular}[c]{@{}c@{}} 0.453 \\ (0.006)\end{tabular}   \\
Time (m) & 15.12 &  15.36 & $>24$ hrs & 20.5 & 21.4 \\ \hline
\end{tabular}
\end{center}

\end{multicols}

\end{document}

%% file: math-commands.tex
\newcommand{\ben}{\begin{enumerate}}
\newcommand{\een}{\end{enumerate}}
\newcommand{\beq}{\begin{equation}}
\newcommand{\eeq}{\end{equation}}
\newcommand{\bde}{\begin{description}}
\newcommand{\ede}{\end{description}}

\newcommand{\argmax}{\operatornamewithlimits{argmax}}

\newcommand{\abs}[1]{\lvert#1\rvert}

\newcommand{\NormRV}{\mathcal{N}}

\newcommand{\CauchyRV}{\mathcal{C}a}

\newtheoremstyle{slplain}
  {1\baselineskip\@plus.2\baselineskip\@minus.2\baselineskip}
  {.5\baselineskip\@plus.2\baselineskip\@minus.2\baselineskip}
  {\slshape}
  {}
  {\bfseries}
  {.}
  { }
  {}

\theoremstyle{slplain}

\newtheorem{theorem}{Theorem}

\newtheorem{corollary}[theorem]{Corollary}

\newtheorem{remark}[theorem]{Remark}

\usepackage{booktabs,array}

\newcount\rowc